\documentclass[useAMS,usenatbib,usegraphicx]{mn2e}

\newcommand{\nc}{\newcommand}
\nc{\aov}{\alpha_\mathrm{ov}}
\nc{\ct}{C_\mathrm{t}}
\nc{\et}{E_\mathrm{t}}
\nc{\etv}{\dot E_\mathrm{tv}}
\nc{\fc}{F_\mathrm{c}}
\nc{\fr}{F_\mathrm{r}}
\nc{\ft}{F_\mathrm{t}}
\nc{\hp}{H_\mathrm{P}}
\nc{\mzams}{M_\mathrm{ZAMS}}
\nc{\pt}{P_\mathrm{t}}
\nc{\teff}{T_\mathrm{eff}}
\nc{\tev}{t_\mathrm{ev}}
\nc{\utv}{\dot U_\mathrm{tv}}

\title[Evolutionary status of Polaris]{Evolutionary status of Polaris}

\author[Yu. A. Fadeyev]{Yu. A. Fadeyev\thanks{E-mail: fadeyev@inasan.ru}\\
Institute of Astronomy of the Russian Academy of Sciences, 48 Pyatnitskaya St., 119017, Moscow, Russia}

\begin{document}

\date{Accepted 2015 February 23. Received 2015 February 21; in original form 2015 February 1}

\pagerange{\pageref{firstpage}--\pageref{lastpage}} \pubyear{}

\maketitle

\label{firstpage}

\begin{abstract}
Hydrodynamic models of short--period Cepheids were computed to determine
the pulsation period as a function of evolutionary time during the first
and third crossings of the instability strip.
The equations of radiation hydrodynamics and turbulent convection for radial
stellar pulsations were solved with the initial conditions obtained from
the evolutionary models of population~I stars ($X=0.7$, $Z=0.02$) with
masses from $5.2M_\odot$ to $6.5M_\odot$ and the convective core overshooting
parameter $0.1\le\aov\le 0.3$.
In Cepheids with period of 4~d the rate of pulsation period change during
the first crossing of the instability strip is over fifty times larger than
that during the third crossing.
Polaris is shown to cross the instability strip for the first time and
to be the fundamental mode pulsator.
The best agreement between the predicted and observed rates of period change
was obtained for the model with mass of $5.4M_\odot$ and the overshooting parameter
$\aov=0.25$.
The bolometric luminosity and radius are $L = 1.26\cdot 10^3L_\odot$ and
$R = 37.5 R_\odot$, respectively.
In the HR diagram Polaris is located at the red edge of the instability strip.

\end{abstract}


\section{Introduction}

Polaris ($\alpha$~UMi) is a short--period ($\Pi\approx 3.969$~d) small--amplitude
Cepheid with steadily increasing pulsation period \citep{f66,af83,dmwh89}.
The rate of period change is estimated from
$\dot\Pi = 4.5\ \mathrm{s\ yr}^{-1}$ \citep{tsdab05} to
$\dot\Pi = 4.9\ \mathrm{s\ yr}^{-1}$ \citep{ss08}.
These values are over an order of magnitude larger than the rates of period change
in most of the Cepheids with a period of $\Pi\approx 4$~d.
This lead \citet{bipt97} to the conclusion that Polaris is undergoing the first crossing
of the instability strip after the main--sequence phase.
\citet{neglww12} and \citet{n14} however casted doubts on this assumption
and argued that Polaris is evolving along the blue loop.
They also showed that the observed rate of period change is more consistent
with predictions from stellar evolution models provided that Polaris is currently
loosing mass at a rate of $\dot M \sim 10^{-6} M_\odot~\mathrm{yr}^{-1}$.

Since about the middle twentieth century Polaris showed decrease
in photometric and radial velocity amplitudes \citep{af83,fks93}.
\citet{dmwh89} supposed that Polaris is approaching the red edge
of the instability strip but in 1994--1997 the decrease in amplitude
has stopped \citep{kf98}.
Recent photometric and radial velocity measurements reveal
slow increase of the pulsation amplitude \citep{besp08,lmhpk08,ss08}.
The cause of the amplitude growth remains uknown.
\citet{fks93}, \citet{ess02}, \citet{tsdab05} reported that Polaris
lies well inside the instability strip.

Polaris is the nearest Cepheid and determination of its pulsation mode is
of great importance in establishing the Cepheid period--luminosity relation.
\citet{fc97} showed that the \textit{Hipparcos} trigonometrical parallax data
are consistent with first overtone pulsation in Polaris.
This conclusion was later confirmed by nonlinear stellar pulsation models by \citet{bgmf01}.
However recent measurements of photometric and spectroscopic parallaxes
seem to be more consistent with fundamental mode pulsation \citep{tkug13}.

The Hipparcos parallax of $7.54 \pm 0.11$~ mas \citep{vanL07} implies a distance
to Polaris of $d=133\pm 2$~ pc but the spectroscopic and photometric parallaxes
give the distance from $d=99 \pm 2$~pc \citep{tkug13} to $d=110$ pc \citep{uk08}.
Therefore the mean angular diameter of 3.28 mas measured with the Navy Prototype Optical
Interferometer \citep{naghhsh00} corresponds to the linear radius of $R=47R_\odot$
for the distance $d=133$~pc and $R=35R_\odot$ for $d=99$~pc.

Polaris is the primary component of a close binary system with an orbital period of 29.6~yr.
\citet{esb08} directly detected the secondary component
using UV images ($\lambda \sim 2255 \AA$) obtained with the Hubble Space Telescope
and found that Polaris has a mass in the range $M=3.1$ -- $6.7 M_\odot$.
This is the first direct estimate of the Cepheid dynamical mass.

In this work, we aim to clarify some contradictions mentioned above.
We calculate the rates of period change in Cepheids with periods near 4~d and
determine the evolutionary status of Polaris by comparing the observed rate
with theoretical predictions.
To this end we compute the non--linear stellar pulsation models as solution
of the equations of radiation hydrodynamics and turbulent convection
with initial conditions obtained from the evolutionary computations.
Such an approach the author used earlier to obtain theoretical rates
of pulsation period change in LMC and Galactic Cepheids \citep{f13,f14}.
In comparison with previous works both evolutionary and hydrodynamical computations
of this study were carried out with improved input physics.
As shown below, determination of the evolutionary status provides constraints on
the pulsation mode and the fundamental parameters of Polaris.

\section{Methods of computation}

\subsection{Stellar evolution}

This work is aimed at determining the pulsation period $\Pi$ as a function
of evolutionary time $\tev$ during the first and third crossings of
the Cepheid instability strip.
We computed the evolutionary tracks for non--rotating population~I stars
with initial masses $5.2 M_\odot\le\mzams\le 6.5 M_\odot$ evolving
from the zero age main sequence (ZAMS) until the end of core helium burning.
Initial mass fractional abundances of hydrogen and heavy elements
are $X=0.7$ and $Z=0.02$, respectively.
The initial relative abundances of the elements heavier than helium
normalized by $Z$ were determined accoring to \citet{agss09}.
Evolutionary computations were done with a code implementing
the Henyey method \citep{f07,f10}.
Below we comment some improvements used in the present work.

Thermodynamic quantities are obtained by interpolation in the tables
calculated with the library of programs FreeEOS \citep{i12}.
Rosseland opacities are computed from OPAL data \citep{ir96}
complemented at low temperatures by the opacities from \citep{faa05}.
The nuclear reaction rates are from the NACRE database \citep{a99}
with updates for reactions ${}^{14}\mathrm{N}(p,\gamma){}^{15}\mathrm{O}$,
${}^4\mathrm{He}(\alpha\alpha,\gamma){}^{12}\mathrm{C}$
and ${}^{12}\mathrm{C}(\alpha,\gamma){}^{16}\mathrm{O}$
that were taken from the REACLIB database \citep{rsstw10}.

In the evolutionary computations convection is treated according to the
mixing length theory \citep{b-v58} using a mixing--length parameter
$\alpha_\Lambda=1.6$.
Boundaries of convective zones are determined with the Schwarzschild criterion
but the convective core is extended in radius by $\aov\hp$, where
$\hp$ is the pressure scale height and $\aov$ is the dimensionless overshooting
parameter ranging from 0.1 to 0.3.

To assure the gradual growth of the convective core during the core helium burning
phase the evolutionary models were computed with $\sim 10^4$ mass zones.
The mass loss rate $\dot M$ during the main sequence phase is from formulae
by\citet{vkl99,vkl00,vkl01} and outside their domain $\dot M$ is evaluated
according to \citet{nj90}.
It should be noted that in our problem the role of mass loss is insignificant.
For example, the mass of the Cepheid $\mzams=5.9M_\odot$ during the
first and third crossings of the instability strip is $M=5.89M_\odot$
and $M=5.84M_\odot$, respectively.

Bearing in mind the importance of initial conditions for correct determination
of the predicted rates of pulsation period change we carried out additional
evolutionary computations for stars with initial composition $X=0.72$, $Z=0.014$
and the overshooting parameter $\aov=0.1$.
Results of these computations were compared with evolutionary models
of non--rotating stars with initial masses $5M_\odot$ and $7M_\odot$ \citep{egem12}.
The difference in luminosity and time scales was found to be less than 5 per cent.

\subsection{Non--linear stellar pulsation}

To compute non--linear stellar pulsation models we solve the Cauchy problem
for the equations of one--dimensional radiation hydrodynamics with
time--dependent turbulent convection treated according to \citet{k86}.
The equations are written as
\begin{equation}
\label{erh1}
\frac{\mathrm{d}U}{\mathrm{d}t} = - \frac{G M_r}{r^2}
 - \frac{1}{\rho} \frac{\partial}{\partial r} \left(P + P_\mathrm{t}\right) + \utv ,
\end{equation}
\begin{equation}
\label{erh2}
\frac{\mathrm{d}E}{\mathrm{d}t} + P \frac{\mathrm{d}V}{\mathrm{d}t} =
 - \frac{1}{\rho r^2} \frac{\partial}{\partial r}\left[r^2\left(\fr + \fc\right)\right] - \ct ,
\end{equation}
\begin{equation}
\label{erh3}
\frac{\mathrm{d}\et}{\mathrm{d}t} + \pt \frac{\mathrm{d}V}{\mathrm{d}t} =
 - \frac{1}{\rho r^2} \frac{\partial}{\partial r}\left(r^2 \ft\right) + \ct + \etv .
\end{equation}
Here $t$ is time, $G$ is the Newtonian constant of gravitation,
$M_r$ is the mass interior to radius $r$,
$U$ is the velocity of the gas flow,
$P$ is the total (gas plus radiation) pressure,
$\rho$ is gas density and $V=1/\rho$ is specific volume,
$E$ is the specific internal energy of gas,
$\fr$ is the radiation flux computed in the diffusion approximation.
The turbulent pressure $\pt$ is related to the specific turbulent kinetic
energy $\et$ by $\pt = \frac{2}{3}\rho \et$.

The convective flux is
\begin{equation}
\fc = \rho T \Pi ,
\end{equation}
where $T$ is the gas temperature,
\begin{equation}
\label{corenvel}
\Pi = - \alpha_\Lambda \alpha_\mathrm{s} \hp \et^{1/2} \frac{\partial S}{\partial r}
    = \alpha_\Lambda \alpha_\mathrm{s} \et^{1/2} C_\mathrm{P} \left(\nabla - \nabla_\mathrm{a}\right)
\end{equation}
is the correlation between entropy and velocity fluctuations,
$\nabla = d\ln T/d\ln P$,
$\nabla_\mathrm{a} = \left(\partial\ln T/\partial\ln P\right)_S$,
$C_\mathrm{P}$ is the specific heat at constant pressure,
$S$ is the specific entropy.

The turbulent kinetic energy flux is
\begin{equation}
\label{fturb}
\ft = - \alpha_\Lambda \alpha_\mathrm{t} \rho \hp \et^{1/2} \frac{\partial\et}{\partial r} .
\end{equation}
The coupling term is
\begin{equation}
\ct = S_\mathrm{t} - D_\mathrm{t} - D_\mathrm{r} ,
\end{equation}
where
\begin{equation}
S_\mathrm{t} = \frac{T}{\hp} \nabla_\mathrm{a} \Pi
\end{equation}
is the source of specific turbulent energy due to buoyancy forces,
\begin{equation}
\label{dismol}
D_\mathrm{t} = \frac{\alpha_\mathrm{d}}{\alpha_\Lambda} \frac{\et^{3/2}}{\hp}
\end{equation}
is the dissipation of specific turbulent energy due to molecular viscosity.
The radiative dissipation is
\begin{equation}
D_\mathrm{r} = \frac{\et}{\tau_\mathrm{rad}} ,
\end{equation}
where
\begin{equation}
\label{taurad}
\tau_\mathrm{rad} = \frac{\alpha_\Lambda^2}{\gamma_\mathrm{rad}^2}\frac{C_\mathrm{P}\kappa P^2}{4\sigma g^2 T^3} ,
\end{equation}
$\sigma$ is the Stefan--Boltzmann constant,
$\kappa$ is opacity,
$g = G M_r/r^2$ is the gravitational acceleration.

Viscous momentum and energy transfer rates are
\begin{equation}
\utv = \frac{1}{r^3\rho} \frac{\partial}{\partial r}
       \left[\frac{4}{3}\mu_\mathrm{tv} r^3 \left(\frac{\partial U}{\partial r} - \frac{U}{r}\right)\right]
\end{equation}
and
\begin{equation}
\etv = \frac{4}{3}\frac{\mu_\mathrm{tv}}{\rho} \left(\frac{\partial U}{\partial r} - \frac{U}{r}\right)^2 ,
\end{equation}
respectively, where
\begin{equation}
\label{mutv}
\mu_\mathrm{tv} = \alpha_\Lambda\alpha_\mathrm{m} \rho \hp \et^{1/2}
\end{equation}
is the kinetic turbulent viscosity.

Formulae (\ref{corenvel}), (\ref{fturb}), (\ref{dismol}), (\ref{taurad}) and (\ref{mutv})
contain dimensionless parameters $\alpha_\Lambda$, $\alpha_\mathrm{s}$, $\alpha_\mathrm{t}$,
$\alpha_\mathrm{d}$, $\gamma_\mathrm{rad}$ and $\alpha_\mathrm{m}$.
The mixing length to pressure scale height ratio is the same as in the evolutionary
computations: $\alpha_\Lambda=1.6$.
Parameters $\alpha_\mathrm{s}$, $\alpha_\mathrm{t}$,
and $\alpha_\mathrm{d}$ are set according to \citet{wf98}:
$$
\alpha_\mathrm{s} = \frac{1}{2}\sqrt{\frac{2}{3}} ,\quad
\alpha_\mathrm{t} = 0.6093 \alpha_\mathrm{s} , \quad
\alpha_\mathrm{d} = \frac{8}{3}\sqrt{\frac{2}{3}} .
$$
Appropriate values $\gamma_\mathrm{rad} = 0.4\sqrt{3}$ and $\alpha_\mathrm{m} = 0.1$
were determined from trial computations of Cepheid models \citep{f14}.
Thus, the parameter $\gamma_\mathrm{rad}$ in the expression for the radiative cooling time of
convective elements (\ref{taurad}) is five times smaller than originally
proposed by \citet{wf98}.
It should be noted that nearly the same parameters were found to be most appropriate
for Cepheid modelling in works by \citet{kbsc02}, \citet{sbb07}, \citet{sm08,sm10}.

In our study we consider the self--excited radial oscillations.
The initial hydrodynamic model is computed by interpolating the zonal
quantities of the evolutionary model and interpolation errors play the
role of small initial perturbations.
To diminish the amplitude of initial perturbations we used fine zoning in mass
in the outer layers of the evolutionary model so that as many as $3\cdot 10^3$
outer mass zones had the relative radius $r > 0.1R$, where $R$ is the radius of the star.
All hydrodynamic models were computed for $N=500$ mass zones.
The mass intervals increase geometrically inward with the ratio of $q\approx 1.03$.
The radius $r_1$ and luminosity $L_1$ at the innermost zone
are assumed to be time--independent.
The radius of the inner boundary satisfies the condition $r_1\le 0.1R$.

\section{Rates of pulsation period change}

Together with numerical integration of equations (\ref{erh1}) -- (\ref{erh3})
with respect to time $t$ we calculate the kinetic energy
\begin{equation}
E_\mathrm{K}(t) = \frac{1}{2}\sum\limits_{j=2}^N \Delta M_{j-1/2} U_j(t)^2 ,
\end{equation}
where $\Delta M_{j-1/2} = M_j - M_{j-1}$ and $M_j$ is the Lagrangian coordinate
of the $j$--th mass zone.
The kinetic energy reaches the maximum value $E_\mathrm{K,max}$ twice per pulsation
period $\Pi$ and for the initial interval of integration with exponential change of
$E_\mathrm{K,max}$ we evaluate the pulsation growth rate
$\eta = \Pi d\ln E_\mathrm{K,max}/dt$.
The period $\Pi$ is determined by a discrete Fourier transform of $E_\mathrm{K}(t)$.
The total sampling interval ranges from $10^3$ to $10^4$ pulsation cycles.

\begin{figure}
\includegraphics[width=84mm]{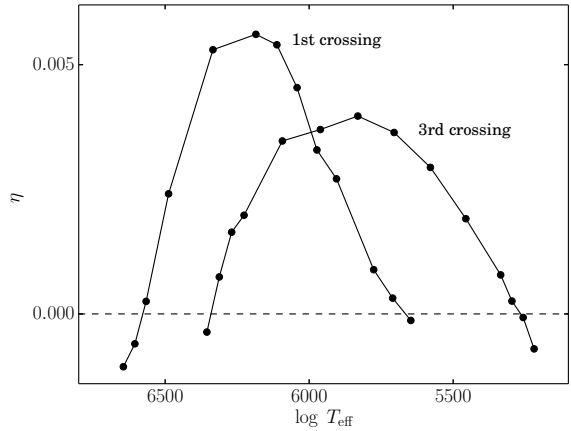}
\caption{The pulsation growth rate $\eta$ as a function of effective temperature $\teff$
         during the first and third crossings of the instability strip for
         the evolutionary sequence $\mzams=5.9M_\odot$, $\aov=0.1$.
         Hydrodynamic models are shown by filled circles.}
\label{fig1}
\end{figure}

Fig.~\ref{fig1} shows the pulsation growth rate $\eta$ as a function of
effective temperature $\teff$ during the first and third crossings
of the instability strip for hydrodynamic models of the evolutionary sequence
$\mzams=5.9M_\odot$, $\aov=0.1$.
The evolutionary time $\tev$, stellar luminosity $L$ and stellar radius $R$
at the edges of the instability strip are determined for $\eta = 0$
by linear interpolation between two adjacent hydrodynamic models with
growth rates of opposite signs.
The time spent in the instability strip is $\Delta t_\mathrm{ev,1} = 1.60\cdot 10^4$~yr
during the first crossing and $\Delta t_\mathrm{ev,3} = 4.26\cdot 10^5$~yr
during the third crossing.
The ratio $\Delta t_\mathrm{ev3}/\Delta t_\mathrm{ev1}$ increases with mass
and ranges from 25 to 30 for $5.6M_\odot\le\mzams\le 6.1M_\odot$.

\begin{figure}
\includegraphics[width=84mm]{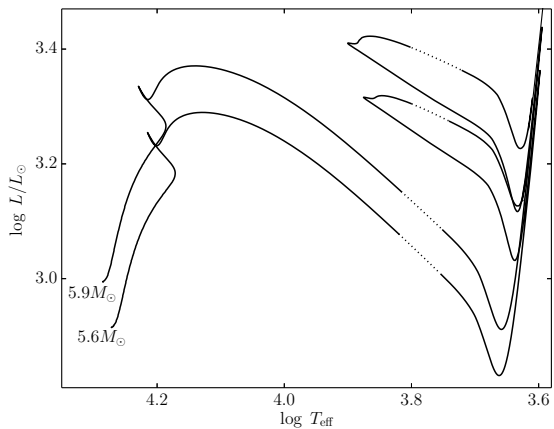}
\caption{Evolutionary tracks of stars $\mzams=5.6M_\odot$ and $\mzams=5.9M_\odot$
         ($\aov=0.1$). The sections of tracks corresponding to the first and third
         crossings of the instability strip are indicated by dotted lines.}
\label{fig2}
\end{figure}

Evolutionary tracks for stars with initial masses $5.6M_\odot$ and $5.9M_\odot$
($\aov=0.1$) are displayed in Fig.~\ref{fig2}.
The sections of the evolutionary tracks corresponding to pulsations with
increasing period (the first and third crossings of the instability strip)
are shown by dotted lines.
Cepheids with decreasing periods are beyond the scope of the present study and
therefore the second crossing of the instability strip is not indicated in the plots.

\begin{figure}
\includegraphics[width=84mm]{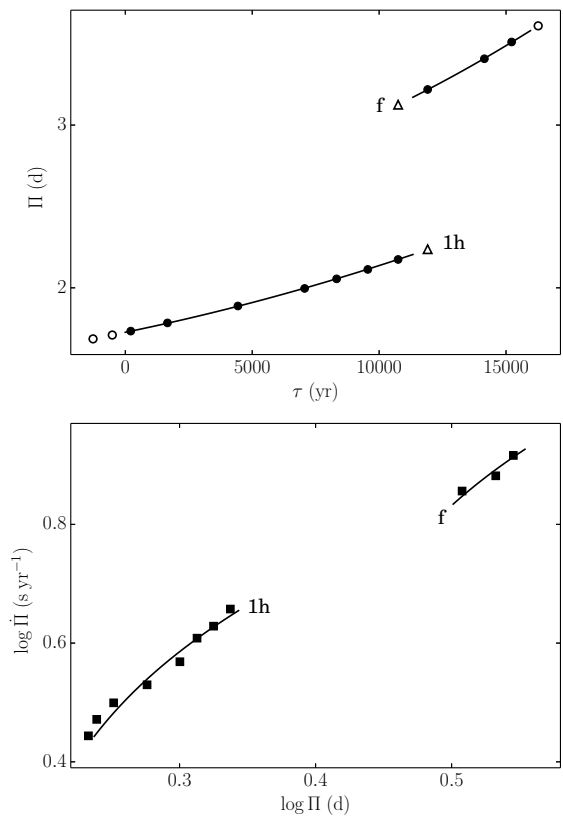}
\caption{The pulsation period $\Pi$ as a function of evolutionary time $\tau$
         (upper panel) and the rate of period change $\dot\Pi$ as a function
         of period $\Pi$ (lower panel) during the first crossing of the instability strip
         for the evolutionary sequence $\mzams=5.9M_\odot$, $\aov=0.1$.
         In the upper panel solid lines show the second--order polynomial fits,
         open circles and filled circles show the periods of pulsationally stable and
         pulsationally unstable models, respectively.
         Open triangles show the periods of the secondary mode near the mode
         switching.
         In the lower panel solid lines show time derivatives of analytical fits.
         Solid squares represent the time derivatives obtained by numerical differentiation.}
\label{fig3}
\end{figure}

The upper panel of Fig.~\ref{fig3} shows the temporal dependence of the
pulsation period $\Pi$ during the first crossing of the instability strip
for the evolutionary sequence $\mzams=5.9M_\odot$, $\aov=0.1$.
For the sake of convenience we use $\tau = \tev-t_\mathrm{ev,b}$
as an independent variable, where $t_\mathrm{ev,b}$ is the evolutionary time
at the blue edge of the instability strip.
Periods of pulsationally stable and pulsationally unstable models are shown
by open circles and filled circles, respectively.

During the first crossing of the instability strip the star begins to pulsate
in the first overtone but as the star approaches the red edge it becomes
the fundamental--mode pulsator.
For hydrodynamic models near the pulsational mode switching we evaluated the periods
of the first overtone (1h) and fundamental mode (f) using the additional
discrete Fourier transform of $E_\mathrm{K}$.
The sampling interval was limited by the amplitude growth stage when the power spectrum
reveals the presence of secondary modes with decaying amplitudes.
Periods of such modes are shown in the upper panel of Fig.~\ref{fig3}
by open triangles.
The boundary between the first overtone and fundamental mode pulsators is defined
as a middle point between two adjacent hydrodynamic models with different
orders of the primary mode.
For the first crossing of the instability strip the second--order algebraic polynomials
approximate the temporal dependence $\Pi(\tau)$ with an rms error less
than 0.1\%.
In the upper panel of Fig.~\ref{fig3} the analytical fits are shown by solid lines.

Time derivatives of analytical fits to $\Pi(\tau)$ are shown in the lower panel
of Fig.~\ref{fig3} as a function of pulsation period $\Pi$.
For the sake of convenience the rate of pulsation period change $\dot\Pi$
is expressed in units of seconds per year.
Solid squares show estimates of $\dot\Pi$ obtained by numerical differentiation.
The small deviation of numerical derivatives from the analitical fits
to $\dot\Pi(\tau)$ is due to the limited accuracy of evolutionary calculations.

\begin{figure}
\includegraphics[width=84mm]{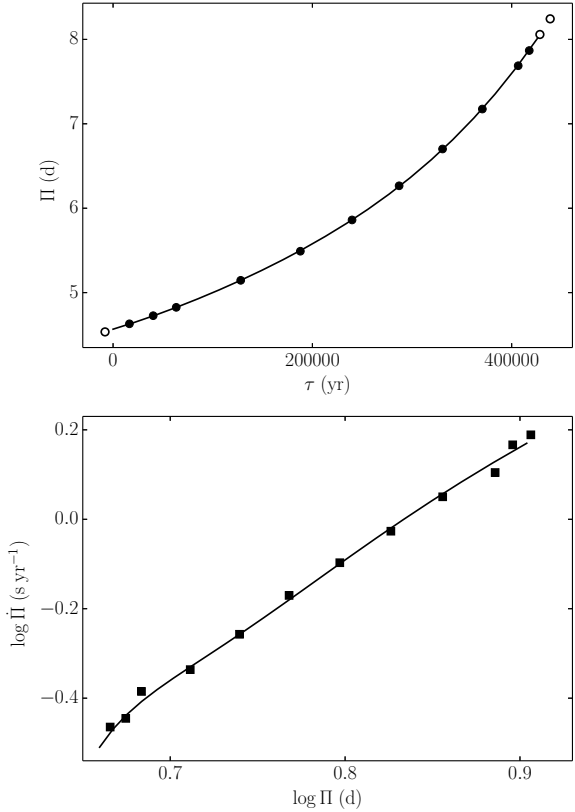}
\caption{Same as Fig.~\ref{fig3}, but for the third crossing of the instability strip.
         All hydrodynamic models show pulsations in the fundamental mode.
         The solid lines represent the fourth--order polynomial fitting (upper panel)
         and its time derivative (lower panel).
}
\label{fig4}
\end{figure}

Fig.~\ref{fig4} displays the temporal dependence of $\Pi$ (upper panel)
and the plot of $\dot\Pi$ as a function of $\Pi$ (lower panel)
for models of the evolutionary sequence $\mzams=5.9M_\odot$, $\aov=0.1$ during the
third crossing of the instability strip.
All hydrodynamic models were found to pulsate in the fundamental mode.
The temporal dependence $\Pi(\tau)$ is fitted with the relative rms error less than 0.1\%
by the fourth--order algebraic polynomial.

\begin{figure}
\includegraphics[width=84mm]{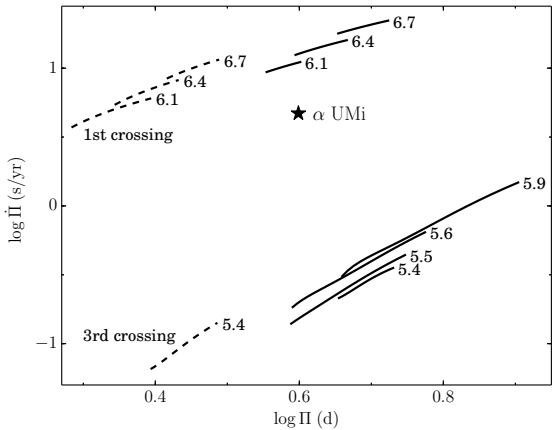}
\caption{The rate of period change $\dot\Pi$ as a function of pulsation period $\Pi$
         for evolutionary sequences with $\aov=0.1$ during the first and
         third crossings of the instability strip.
         The numbers at the curves indicate the initial stellar mass $\mzams$
         in solar units.
         Solid and dashed lines correspond to the fundamental mode and first overtone,
         respectively.
}
\label{fig5}
\end{figure}

The results of polynomial fitting for hydrodynamic models with pulsation periods
$2~\mathrm{d}\le\Pi\le 8~\mathrm{d}$
are displayed in Fig.~\ref{fig5} where the rates of period change $\dot\Pi$ are
plotted as a function of period $\Pi$ for evolutionary sequences with initial masses
$5.4 M_\odot\le\mzams\le 5.9 M_\odot$ and the overshooting parameter $\aov=0.1$.
Each curve in this diagram describes the evolution of values $\Pi$, $\dot\Pi$
as the star crosses the instability strip.
The solid and dashed lines represent the fundamental mode and first overtone,
respectively.
The plots located in the upper part of Fig.~\ref{fig5} represent the first crossing
of the instability strip and the plots located in the lower part of the diagram
correspond to the third crossing.
As is clearly seen, for Cepheids with a period of $\Pi\approx 4$~d
the rate of period change during the first crossing of the instability strip
$\dot\Pi_1$ is over fifty times larger than the rate of period change during
the third crossing $\dot\Pi_3$.
It should be noted that large values of the ratio $\dot\Pi_1/\dot\Pi_3$
are also typical for long--period Cepheids \citep{f14}.

The position of Polaris is marked in Fig.~\ref{fig5} for the period $\Pi_\star=3.969$~d
and the rate of period change $\dot\Pi_\star=4.7~\mathrm{s~yr}^{-1}$
which is the mean value of the observational estimates by
\citet{tsdab05} and \citet{ss08}.
Even a cursory glance at Fig.~\ref{fig5} leads us to the following conclusions:
(1) Polaris crosses the instability strip for the first time;
(2) Polaris is the fundamental mode pulsator.
Below we discuss in more detail both these conclusions.

\begin{figure}
\includegraphics[width=84mm]{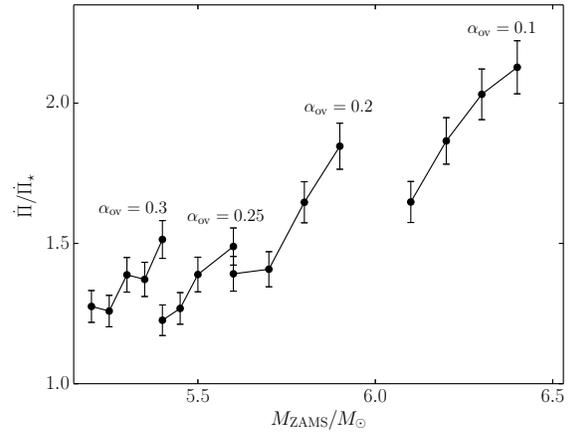}
\caption{The ratio of the predicted to observed rates of period change
         $\dot\Pi/\dot\Pi_\star$
         as a function of initial stellar mass $\mzams$ for
         the overshooting parameters $\aov=0.1$, 0.2, 0.25 and 0.3.
         }
\label{fig6}
\end{figure}

The diagram in Fig.~\ref{fig5} was obtained for the evolutionary sequences
computed with the overshooting parameter $\aov=0.1$ and the ratio
of the predicted to observed rates of period change is $\dot\Pi/\dot\Pi_\star\approx 2$.
However the rate of period change during the first crossing of the instability strip
depends on the assumed overshooting during the main--sequence evolutionary phase,
so that agreement between theory and observations can be substantially improved.
This is illustrated by Fig.~\ref{fig6} where the ratio $\dot\Pi/\dot\Pi_\star$
is plotted as a function of initial mass $\mzams$ for several values
of the overshooting parameter $\aov$.
The upper and lower errorbars correspond to the observational estimates
by \citet{tsdab05} and \citet{ss08}, respectively.
As seen from Fig.~\ref{fig6}, the best agreement between the predicted and
observed rates of period change ($\dot\Pi/\dot\Pi_\star\approx 1.25$)
is obtained for the overshooting parameter $\aov=0.25$ and the
initial stellar mass $5.4M_\odot\le\mzams\le 5.45M_\odot$.

\section{Fundamental parameters of Polaris}

\begin{figure}
\includegraphics[width=84mm]{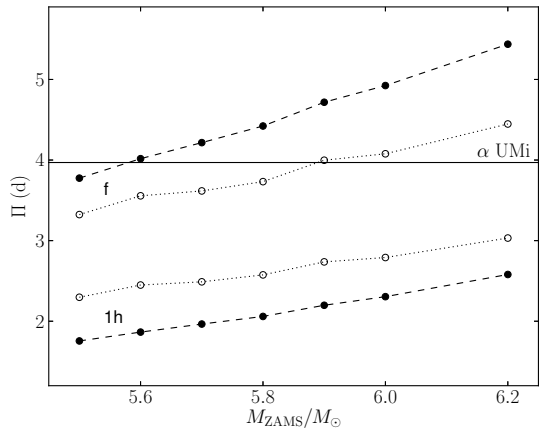}
\caption{The pulsation period at the blue edge (lower dashed line) and
         red edge (upper dashed line) of the instability strip
         for models of first--crossing Cepheids computed with
         the overshooting parameter $\aov=0.2$.
         Lower and upper dotted lines show the periods of the first
         overtone and fundamental mode at the mode switch boundary.
         Filled and open circles represent the interpolated values of the period.
         }
\label{fig7}
\end{figure}

To determine the pulsation mode of first--crossing Cepheids with period of $\Pi=3.969$~d
we considered the grid of evolutionary and hydrodynamic models
with initial stellar masses $5.2M_\odot\le\mzams\le 6.5M_\odot$
and the overshooting parameter ranging from 0.1 to 0.3.
Typical results obtained for the overshooting parameter $\aov=0.2$ are shown
in Fig.~\ref{fig7}.

The filled circles connected by dashed lines in Fig.~\ref{fig7} show the pulsation
period at the blue and red edges of the instability strip, whereas
the open circles connected by dotted lines show the pulsation period
at the mode switch.
Thus, for a fixed value of $\mzams$ the plots in Fig.~\ref{fig7}
show the range of periods for pulsations in the first overtone
and in the fundamental mode.
The similar diagrams were obtained for other values of the overshooting parameter
$\aov$ and no models pulsating in the first overtone with period of $\Pi=3.969$~d
were found.
Therefore, Polaris is undoubtedly a fundamental--mode pulsator.

Intersection of the dashed and dotted lines with the solid horizontal line
$\Pi=3.969$~d in Fig.~\ref{fig7} gives the lower and upper values for
the initial stellar mass $\mzams$, respectively.
The upper and lower values of the mass $M$, radius $R$ and luminosity $L$
are evaluated by interpolation between the evolutionary tracks.

\begin{figure}
\includegraphics[width=84mm]{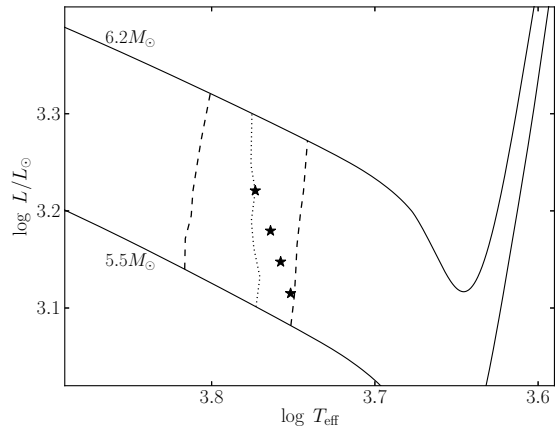}
\caption{Evolutionary tracks of stars with initial masses $5.5M_\odot$ and $6.2M_\odot$
         ($\aov=0.2$) near the first crossing of the instability strip.
         Dashed lines show the blue and red edges of the instability strip.
         The dotted line shows the mode switch boundary.
         The star symbols show the location of models with 
         the pulsation period $\Pi_\star=3.969$~day.
         }
\label{fig8}
\end{figure}

Fig.~\ref{fig8} shows the evolutionary tracks in the HR diagram for stars
with initial masses $5.5M_\odot$ and $6.2M_\odot$ ($\aov=0.2$).
The dashed lines show the blue and red edges of the instability strip.
The mode switch boundary shown by the dotted line is not the straight line
due to difficulties in determining the exact location of the point of mode switch.
Models with pulsation period $\Pi_\star=3.969$~day are shown by the star symbol.

The lower and upper values of fundamental parameters for Polaris are given
in Table~\ref{fpar}.
The last column gives the distance to Polaris $d$ evaluated from the
observed angular diameter $\theta_L = 3.28$~mas \citep{naghhsh00}.

\begin{table}
\caption{The upper and lower values of the fundamental parameters of Polaris}
\label{fpar}
\begin{tabular}{@{}lccccc}
\hline
 $\aov$ & $M/M_\odot$ & $L/L_\odot$ & $R/R_\odot$ & $\teff$ &  $d$  \\
        &             &             &             &  (K)    &  (pc) \\
\hline
  0.1   &       6.38       &  1770       &  39.9       &  5930   &   113  \\
        &       6.08       &  1390       &  39.3       &  5630   &   111  \\
  0.2   &       5.88       &  1660       &  38.7       &  5930   &   110  \\
        &       5.59       &  1300       &  37.8       &  5640   &   107  \\
  0.25  &       5.59       &  1510       &  37.9       &  5850   &   107  \\
        &       5.39       &  1260       &  37.4       &  5640   &   106  \\
  0.3   &       5.39       &  1450       &  37.5       &  5820   &   106  \\
        &       5.19       &  1220       &  36.7       &  5630   &   104  \\
\hline
\end{tabular}
\end{table}

\medskip

\section{Conclusion}

The evolutionary and pulsational models together with the observed period
$\Pi_\star$ allow us to obtain the lower and upper estimates of the initial
mass of Polaris.
As seen in Fig.~\ref{fig6} a major cause of uncertainty in the determination of mass
is the overshooting parameter $\aov$ which is the unknown a priori quantity.
The comparison of theoretical predictions with the observed rate of period increase
leads to the narrower mass range $5.2M_\odot\le\mzams\le 5.7M_\odot$
and to the constraint on the overshooting parameter $\aov\ge 0.2$.
The best agreement with observations ($\dot\Pi/\dot\Pi_\star\approx 1.25$)
was found for the model $\mzams=5.4M_\odot$, $\aov=0.25$ which locates near
the red edge of the instability strip.
The mass, $M=5.39M_\odot$, is within the range
$3.1M_\odot\le M\le 6.7M_\odot$ of the measured dynamical mass \citep{esb08}.

The mean linear radius $R=37.5~R_\odot$ for the overshooting parameter $\aov=0.25$
(see Table~\ref{fpar}) and the observed angular diameter $\theta_L = 3.28$~mas
\citep{naghhsh00} lead to the distance of $d=106$~pc.
This value is consistent with photometric and spectroscopic parallaxes
implying a distance from 99~pc \citep{tkug13} to 110~pc \citep{uk08}.
To further improve theoretical predictions one has to take into account effects
of initial chemical composition.
For example, preliminary computations showed that the increase of the fractional
mass abundance of hydrogen from $X=0.7$ to $X=0.73$ leads to decrease in
$\dot\Pi$ by $\approx 12\%$.

The fact that Polaris is evolving on a Kelvin--Helmholtz timescale after the main
sequence phase and is crossing the instability strip for the first time
allows us to conclude that effects of mass loss do not play
a significant role in evolutionary changes of the pulsation period.

Decrease of the ratio $\dot\Pi/\dot\Pi_\star$ with decreasing initial mass $\mzams$
at the fixed value of the overshooting parameter favours the closer location of Polaris
to the red edge of the instability strip.
This assumption does not necessarily contradict the growth of the pulsation amplitude
observed for two last decades.
Indeed, the balance between the $\kappa$--mechanism exciting pulsational instability
and convection which suppresses oscillations can cyclically change due to
variations of magnetic field on time scale longer than the pulsation period.
Therefore, the currently observed increase of the pulsation amplitude might be
due to the growth of a magnetic field which tends to weaken convection \citep{s09}.
Longitudinal magnetic fields from 10 to 30 Gauss were detected in Polaris by \citep{bfp81},
whereas \citet{ubbp10} reported on detection of variable magnetic field
from -109 to 162 Gauss.

The present study is based on the convection model by \citet{k86}
which contain several parameters of order unity.
The most uncertain of them are $\gamma_\mathrm{rad}$ and  $\alpha_\mathrm{m}$
in expressions for the radiative cooling time (\ref{taurad}) and
kinetic turbulent viscosity (\ref{mutv}), respectively.
The values of these parameters affect the damping of oscillations and
are inferred from observational constraints (for example, the width of the
instability strip).
At the same time, the uncertainty in the values of $\gamma_\mathrm{rad}$ and
$\alpha_\mathrm{m}$ does not play an important role in determinating the
pulsation period.
This is due to the fact that the depth of the convective zone in Cepheids
is only a few per cent of the stellar radius,
whereas the length of the fundamental mode period is of the order of two
sound travel times between the centre and the surface of the star.

\label{lastpage}

\end{document}